\documentstyle[epsf,twoside,fleqn,espcrc2]{article}

\title{In search of a scaling scalar glueball}

\author{Colin Morningstar\address{
           Dept.~of Physics, University of California at San Diego,
           La Jolla, California 92093-0319, USA}
        and 
        Mike Peardon\address{
           HLRZ, Forschungszentrum J\"ulich, J\"ulich D-52425, Germany}
        \thanks{Based on the talk presented by M.P.}
        }
       
\begin{document}

\begin{abstract}
Anisotropic lattices are an efficient means of studying the glueballs of QCD, 
however problems arise with simulations of the lightest, scalar state. The 
mass is strongly dependent on the lattice spacing, even when a mean-field 
improved gluon action is used. The nature and cause of these errors are 
discussed and the scaling properties of the scalar from different lattice 
actions are presented. 
\end{abstract}

\maketitle

\section{THE SCALAR DIP}

Efficient studies of the QCD glueball spectrum are made possible 
on coarse spatial lattices, provided the temporal discretisation is finer.
This permits sufficient measurements of the glueball correlator before
statistical fluctuations dominate \cite{longglue}. 
These studies found good scaling for the $2^{++}$
and $1^{+-}$ glueballs. However for the lightest, scalar state, 
the mass (in units of $r_0$) was seen to fall rapidly to a minimum when the 
spatial lattice spacing is $\approx 0.25$ fm, where the scaling violations are 
$\approx 25\%$ and then rise as the lattice spacing is increased further; the 
``scalar dip''. This effect was also seen in $SU(2)$ 
simulations \cite{su2_dip}, although the dip is less pronounced.

\begin{figure}[t]
\setlength{\epsfxsize}{7.5cm}
\epsfbox{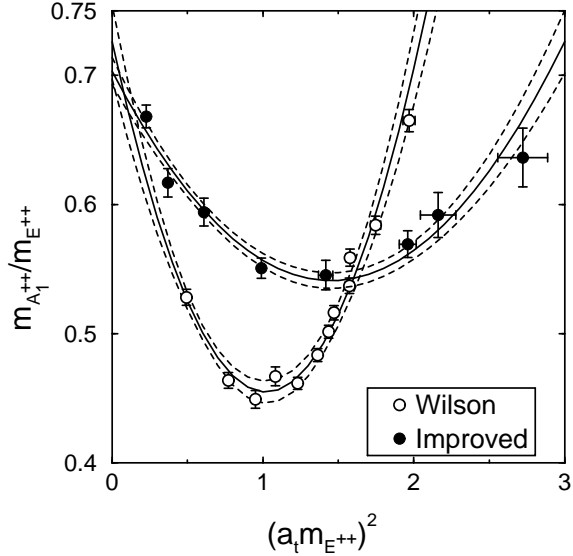}
\vspace{-5.5ex}
\caption{
The effect of Symanzik improvement on the scalar glueball. 
   \label{fig:scaling_1}}
\end{figure}

The first question is whether Symanzik improvement of the action
has worked. Fig. \ref{fig:scaling_1} shows the
scaling of the scalar from both Wilson and improved simulations. In this
figure, the mass of the $E$-irrep tensor glueball is used to set the scale as
these mass ratios are independent of any anisotropy renormalisations. The two 
curves are results of fitting the datasets to a parabola. Fig.
\ref{fig:scaling_1} demonstrates the improvement reduces the depth of the dip,
but only by about a third. 

We are examining a number of other explanations for this behaviour; 
the first is that using the
plaquette definition for the mean-link parameters leads to an under-estimate of
the effects of the tadpole graphs. Secondly, the influence of a critical
end-point in a higher-dimensional space of couplings may be at fault.

In our preliminary study of the mean-link definition no significant 
improvement is observed. This is in contrast to findings for $SU(2)$ where the 
weaker scalar dip was reduced further by employing the Landau-link definition 
\cite{su2_dip}.

Simulations of QCD in the fundamental-adjoint space of plaquette couplings
demonstrate the existence of a critical point close to the fundamental axis.
As this point is approached, the mass of the scalar (in units of the string 
tension) has been shown to fall rapidly \cite{crit_point}.  At the critical
point, at least one lattice correlation length must diverge and this is seen
to be the scalar channel.
If this effect is responsible, addition of
negatively coupled adjoint terms to the action might define a scaling 
trajectory where the influence of this critical point is diminished
\cite{adjt_scalar}.

This solution ought to be 
independent of any Symanzik-improvement programme. To test this, we have
added adjoint-like terms to the anisotropic Wilson action. 
This report discusses progress with these studies. 
Simulations of Symanzik-improved actions with these new terms are in progress. 

\section{DESIGNING NEW ACTIONS}

Based on the observation of the nearby critical point, we consider one member of
a class of new actions which include ``adjoint-like'' terms. The term we
add is designed to be easy to simulate with existing update methods. 

Beginning with the plaquette operator,
\[
P_{\!\mu\nu}(x) = \frac{1}{N} \mbox{ReTr } U_\mu(x) U_\nu(x\!+\!\hat{\mu}) 
                   U^\dagger_\mu(x\!+\!\hat{\nu}) U^\dagger_\nu(x)
\]
the Wilson discretisations of the magnetic and electric field strengths are
\[
\Omega_s = \sum_{x,i>j} \left\{ 1 - P_{ij}(x) \right\} = 
    \frac{\xi}{\beta} \int\!\! d^4\!x \mbox{ Tr } B^2 + {\cal O}(a^2) 
\]
and 
\[
\Omega_t = \sum_{x,i} \left\{ 1 - P_{i0}(x) \right\} = 
    \frac{1}{\xi\beta}\int\!\! d^4\!x \mbox{ Tr } E^2 + {\cal O}(a^2)
\]
respectively, where $i,j$ are spatial indices and $\xi$ is the tree-level
anisotropy, $a_s/a_t$. 
We introduce a term which correlates pairs of spatial plaquettes separated by 
one site temporally 
\[
\Omega_{s}^{(2t)} = \frac{1}{2}\sum_{x,i>j} 
           \left\{ 1 - P_{ij}(x) P_{ij}(x+\hat{t})\right\}.
\]
A similar term with spatially correlated plaquette pairs can be constructed but 
is not considered here.

The separation of the two plaquettes allows the standard Cabibbo-Marinari and 
over-relaxation gauge-field update methods to be applied. Including
two-plaquette terms adds a computational overhead of just $10\%$ to our 
anisotropic Wilson action workstation codes. 

It can be shown that the operator combination
\[
\tilde{\Omega}_s = (1+\omega) \; \Omega_s - \omega \; \Omega_{s}^{(2t)}
\]
has an identical expansion in powers of $a_{s,t}$ to $\Omega_s$ up to 
${\cal O}(a_{s,t}^4)$. Thus a 
signal for being closer to the QCD fixed point would be physical
ratios becoming weakly dependent on the free parameter, $\omega$.

A candidate action 
(without Symanzik improvement) 
is then
\[
S_{\omega} = \beta \left( \xi \; \Omega_t + 
     \frac{1}{\xi} \; \tilde{\Omega}_s \right).
\]
Mean-link improvement is applied in the standard manner.

\begin{figure}[t]
\setlength{\epsfxsize}{7.5cm}
\epsfbox{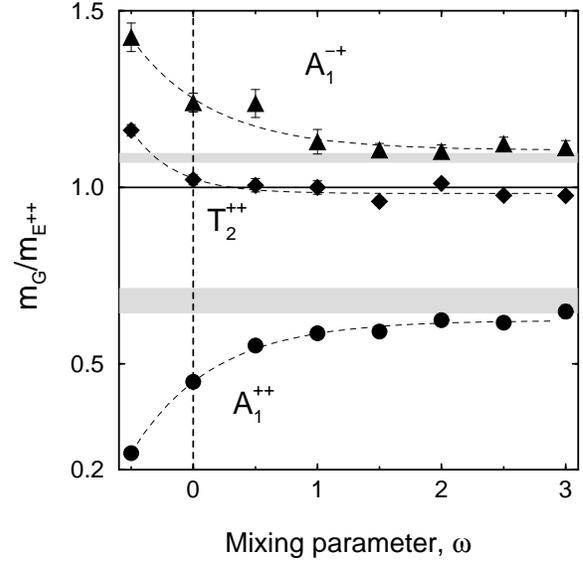}
\vspace{-6ex}
\caption{Glueball mass ratio dependence on the two-plaquette mixing parameter
$\omega$ for the $A_1^{++}, T_2^{++}$ and $A_1^{-+}$ irreps 
at fixed $\beta$. The 
horizontal bands indicate the continuum limit. Dashed lines guide the eye.
   \label{fig:omega}}
\end{figure}

\subsection{Simulation results}

We have performed a range of simulations at a number of values of 
$\omega$ and lattice spacings. 

Fig. \ref{fig:omega} shows the dependence of ratios of
glueball masses on $\omega$ at fixed $\beta$ ($=2.7$) for $S_\omega$.
Close to $\omega=0$ (anisotropic Wilson),
these scaling ratios show a strong dependence on $\omega$, both in the scalar
and pseudoscalar channels. In the range $\omega=1$ to 3, the dependence is much
weaker for all three channels, and the ratios are closer to their continuum 
values. 
The lattice spacing does not change significantly over this range of $\omega$; 
the lattice mass of the $E$ irrep varies by just $7\%$.

\begin{figure}[t]
\setlength{\epsfxsize}{7.5cm}
\epsfbox{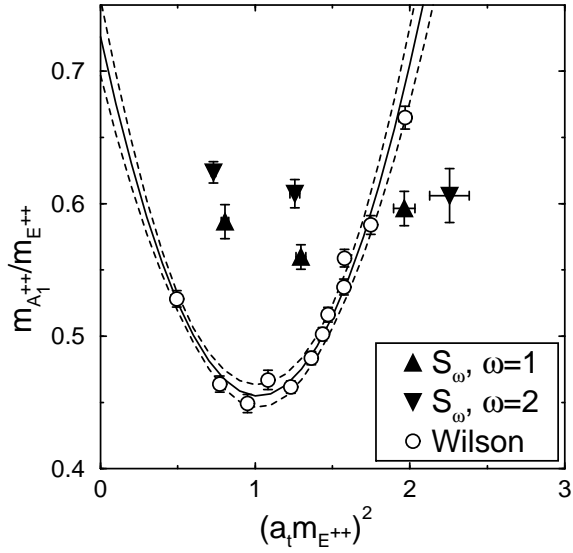}
\vspace{-5ex}
\caption{Scaling properties of the scalar of $S_\omega$.
    \label{fig:scaling_2}}
\end{figure}

Fig. \ref{fig:scaling_2} compares the scaling properties of $S_\omega$
for two values of $\omega$ to the anisotropic Wilson action. 
The cut-off dependence
in the scalar channel is seen to be reduced significantly. More data are
required to resolve the continuum limit fully.

\begin{figure}[t]
\setlength{\epsfxsize}{7.5cm}
\epsfbox{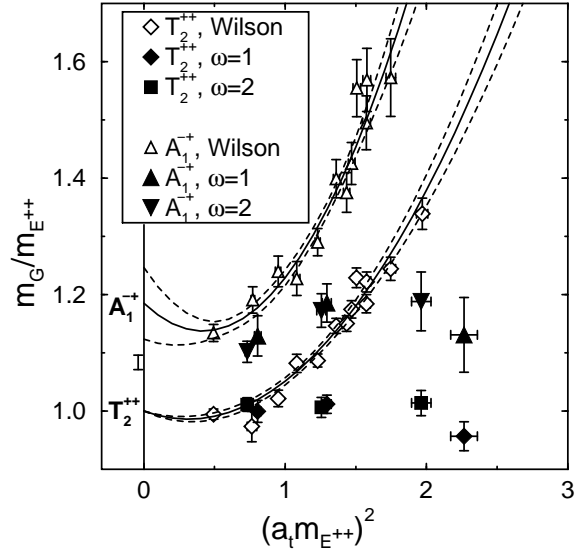}
\vspace{-6.1ex}
\caption{Scaling properties of the $T_2^{++}$ and $A_1^{-+}$ irreps.
Filled symbols indicate $S_\omega$ data. 
\vspace{-0.33cm}
    \label{fig:scaling_3}}
\end{figure}

Fig. \ref{fig:scaling_3} shows the tensor irrep rotational symmetry violations
and the scaling of the pseudoscalar glueball. 
Remarkably, the large scaling violations of the Wilson action in these channels 
are seen almost to vanish for $S_\omega$ with $\omega=1$ or 2. This was not 
anticipated from critical point arguments. 

\section{CONCLUSIONS}

By comparing data from anisotropic Wilson simulations to their Symanzik-improved
counterparts, we have shown that improvement reduces the
scaling violations of the scalar glueball, however large effects remain.

We found that adding to the action a new term which includes the product of two
plaquette traces leads to significant reduction in the curious 
finite-lattice-spacing effects seen for the scalar glueball mass. 
Surprisingly, the
pseudoscalar and $T_2$ tensor irrep masses (in units of the E-irrep mass)  
also show far less cut-off dependence. We emphasise that 
the action, $S_\omega$ used here is not Symanzik improved, and has leading 
discretisation errors at ${\cal O}(a_{s,t}^2)$.

We are performing simulations where two-plaquette terms are
added to Symanzik-improved actions of the type considered in Ref. 
\cite{longglue}.

\end{document}